\documentclass[final]{elsart}
\usepackage{natbib}
\usepackage{graphicx}
\usepackage{amsmath,amssymb}
\journal{Fluid Dynamics Research}
\renewcommand{\cite}{\citep}

\begin{document}
\begin{frontmatter}

\title{Vortex Tubes in Turbulence Velocity Fields at High Reynolds Numbers}
\author{Hideaki Mouri\corauthref{cor1}}
\corauth[cor1]{Corresponding author. {\it E-mail address:} hmouri@mri-jma.go.jp}
\author{and}
\author{Akihiro Hori\corauthref{cor2}}
\corauth[cor2]{Affiliated with Meteorological and Environmental Sensing Technology, Inc., Nanpeidai, Ami 300-0312, Japan}
\address{
Meteorological Research Institute, Nagamine, Tsukuba 305-0052, Japan
}

\begin{abstract}
The elementary structures of turbulence, i.e., vortex tubes, are studied using velocity data obtained in laboratory experiments for boundary layers  and duct flows at microscale Reynolds numbers $Re_{\lambda} = 332$--1934. While past experimental studies focused on intense vortex tubes, the present study focuses on all vortex tubes with various intensities. We obtain the mean velocity profile. The radius scales with the Kolmogorov length. The circulation velocity scales with the Kolmogorov velocity, in contrast to the case of intense vortex tubes alone where the circulation velocity scales with the rms velocity fluctuation. Since these scaling laws are independent of the configuration for turbulence production, they appear to be universal at high Reynolds numbers.
\end{abstract}

\end{frontmatter}

\section{Introduction} \label{s1}

Turbulence contains vortex tubes as the elementary structures \cite{f95,sa97,kida}. Regions of significant vorticity tend to be organized into tubes. They occupy a small fraction of the volume and are embedded in the background fluctuation. Their existence was established at microscale Reynolds numbers $Re_{\lambda} \lesssim 2000$, by seeding a turbulent liquid with gas bubbles and thereby visualizing regions of low pressure that are associated with vorticity \cite{d91,lvmb00}.

At low Reynolds numbers, $Re_{\lambda} \lesssim 200$, direct numerical simulations derived basic parameters of vortex tubes. The radii are of the order of the Kolmogorov length $\eta$. The total lengths are of the order of the correlation length $L$. The circulation velocities are of the order of the Kolmogorov velocity $u_K$ or the rms velocity fluctuation $\langle u^2 \rangle ^{1/2}$. Here $\langle \cdot \rangle$ denotes an average. The lifetimes are of the order of the turnover time for energy-containing eddies $L/ \langle u^2 \rangle ^{1/2}$ \cite{vm91,vm94,j93,jw98,kida,tkmsm04}. 

However, for most of the tube parameters, universality has not been established because the behavior has not been known at high Reynolds numbers. At $Re_{\lambda} \gtrsim 200$, a direct numerical simulation is not easy for now. The bubble visualization does not have a high enough spatial resolution \cite{lvmb00}, except for the study of the tube length and lifetime.

The more promising approach is velocimetry in laboratory experiments. A probe suspended in the flow is used to obtain a one-dimensional cut of the velocity field. The small-scale velocity variation is enhanced at the positions of vortex tubes.\footnote{
Other vortical structures, i.e., vortex sheets, are not important at least statistically. This is already known for the case of intense velocity variation \cite{n97,mhk07} and is to be demonstrated here for the case of velocity variation with arbitrary intensity (\S \ref{s4}).} 
In particular, the velocity component $v$ that is perpendicular to the one-dimensional cut is suited to detecting circulation flows associated with vortex tubes \cite{mtk99}.

Based on this approach, there were already several studies \cite{b96,n97,cg99}. Recently, using boundary layers at $Re_{\lambda} = 332$--1304 and duct flows at $Re_{\lambda} = 719$--1934, we studied the mean radius $R_0$ and circulation velocity $V_0$ of vortex tubes and obtained the scalings $R_0 \propto \eta$ and $V_0 \propto \langle v^2 \rangle ^{1/2}$ \cite{mhk07}. However, this and other past studies focused on intense vortex tubes that are easily captured by imposing a threshold on the velocity variation. For all vortex tubes with various intensities, the tube parameters have not been known. Now we try to study vortex tubes as a whole.

\section{Experimental Data} \label{s2}

The present study is based on data of our past experiments described in \citet{mhk07}. Since wide ranges of the Reynolds number $Re_{\lambda}$ were obtained in two configurations for turbulence production, i.e., boundary layer and duct flow, we are able to study dependence of tube parameters on the Reynolds number and on the large-scale flow. Since the data were long, $(1$--$4) \times 10^8$ points, their statistics are expected to be significant. Table \ref{t1} lists turbulence parameters that are to be used here.

The experiments were done in a wind tunnel of the Meteorological Research Institute. We use coordinates $x$, $y$, and $z$ in the streamwise, spanwise, and floor-normal directions. The origin $x = y = z = 0$~m is on the tunnel floor at the entrance to the test section. Its size was $\delta x = 18$\,m, $\delta y = 3$\,m, and $\delta z = 2$\,m. We simultaneously measured the velocity fluctuations $u$ and $v$ in the streamwise and spanwise directions, by using a hot-wire anemometer with a crossed-wire probe. The wires were 5\,$\mu$m in diameter, 1.25\,mm in sensing length, 1.4\,mm in separation, and oriented at $\pm 45 ^{\circ}$ to the streamwise direction. Taylor's frozen-flow hypothesis was used to convert temporal variations into spatial variations. The sampling interval $\delta x_s$ was set to be as small as possible, on the condition that high-wave-number noise was not significant in the power spectrum.

\begin{table}[t]
\begin{center}
\caption{Turbulence parameters in boundary layers (B1--B6) and duct flows (D1--D5): sampling interval $\delta x_s$, Kolmogorov length $\eta = (\nu ^3 / \langle \varepsilon \rangle )^{1/4}$  where $\nu$ is the kinematic viscosity and $\langle \varepsilon \rangle = 15 \nu \langle (\partial _x v)^2 \rangle /2$ is the mean energy dissipation rate, Taylor microscale $\lambda = [2 \langle v^2 \rangle / \langle (\partial _x v)^2 \rangle ]^{1/2}$, Kolmogorov velocity $u_K = ( \nu \langle \varepsilon \rangle )^{1/4}$, rms velocity fluctuations $\langle u^2 \rangle ^{1/2}$ and $\langle v^2 \rangle ^{1/2}$, and microscale Reynolds number $Re_{\lambda} = \lambda \langle v^2 \rangle ^{1/2} / \nu$. The parameter values are from \citet{mhk07}. \label{t1}}
\vspace*{0.4cm}
\begin{tabular}{cccccccc}
\hline
Data   &${\displaystyle \delta x_s                  \atop[{\rm cm}]              }$
       &${\displaystyle \eta                        \atop[{\rm cm}]              }$
       &${\displaystyle \lambda                     \atop[{\rm cm}]              }$
       &${\displaystyle u_K                         \atop[{\rm m}/{\rm s}]       }$
       &${\displaystyle \langle u^2 \rangle ^{1/2}  \atop[{\rm m}/{\rm s}]       }$
       &${\displaystyle \langle v^2 \rangle ^{1/2}  \atop[{\rm m}/{\rm s}]       }$
       &${\displaystyle Re_{\lambda}                                             }$\\
\hline
B1 & 0.0378 & 0.0539 & 1.93 &   0.0262 & 0.283 & 0.242 & 332 \\
B2 & 0.0312 & 0.0335 & 1.46 &   0.0423 & 0.582 & 0.475 & 488 \\
B3 & 0.0265 & 0.0198 & 1.04 &   0.0716 & 1.18  & 0.973 & 716 \\
B4 & 0.0230 & 0.0152 & 0.919&   0.0934 & 1.80  & 1.46  & 945 \\
B5 & 0.0180 & 0.0120 & 0.776&   0.118  & 2.46  & 1.98  & 1080\\
B6 & 0.0184 & 0.0104 & 0.742&   0.137  & 3.14  & 2.51  & 1304\\
D1 & 0.0355 & 0.0288 & 1.52&   0.0489 & 0.694 & 0.666  & 719 \\
D2 & 0.0255 & 0.0177 & 1.15 &   0.0798 & 1.38  & 1.34  & 1098\\
D3 & 0.0217 & 0.0133 & 0.986&   0.107  & 2.11  & 2.04  & 1416\\
D4 & 0.0216 & 0.0111 & 0.895&   0.128  & 2.84  & 2.69  & 1693\\
D5 & 0.0212 & 0.00955& 0.826&   0.149  & 3.46  & 3.32  & 1934\\
\hline
\end{tabular}
\end{center}
\vspace{0.7cm}
\end{table}

\subsection{Boundary Layers (Data B1--B6)}

Over the entire floor of the tunnel test section, we placed blocks as roughness elements. Their size was $\delta x = 0.06$ m, $\delta y = 0.21$ m, and $\delta z = 0.11$ m. Their spacing was $\delta x = \delta y = 0.5$ m. The measurement position was at $x = 12.5$ m, where the boundary layer had been well developed, and $z = 0.25$--0.35\,m in the log-law sublayer.\footnote{
The measurement position might appear to have been too close to the floor of the tunnel, but this is not serious at all. First, this is not in contradiction to our purpose, which is to find features of vortex tubes that are independent of the large-scale flow. Second, not invalid is the basic assumption of our study that the spatial distribution of vortex tubes was similar among experiments (\S \ref{s3}). Since the measured ratio $\langle u^2 \rangle / \langle v^2 \rangle$ is not far from unity (Table \ref{t1}), the tube distribution at the measurement position should have been almost isotropic in all the experiments.}  
We obtained the data B1--B6 at $Re_{\lambda} = 332$--$1304$ by changing the incoming-flow velocity from 2 to 20\,m\,s$^{-1}$.

\subsection{Duct Flows (Data D1--D5)}

At $x = -2$\,m, we placed a rectangular duct with width $\delta y = 1.3$\,m and $\delta z = 1.4$\,m. The duct center was on the tunnel axis. The measurement position was at $x = 15.5$\,m and $z = 0.6$\,m, where the flow had become turbulent. We obtained the data D1--D5 at $Re_{\lambda} = 719$--$1934$ by changing the duct-exit flow velocity from 11 to 55\,m\,s$^{-1}$.

\section{Model for Vortex Tubes} \label{s3}

Using the Burgers vortex, an idealized model for vortex tubes, we discuss what information is available from a one-dimensional cut of the velocity field (for a similar discussion based on a direct numerical simulation, see \citet{mtk99}). The Burgers vortex is an axisymmetric steady circulation in a strain field. In cylindrical coordinates, the circulation $u_{\Theta}$ and strain field $(u_R, u_Z)$ are
\begin{equation}
u_{\Theta} \propto \frac{\nu}{a_0 R} \left[ 1 - \exp \left( - \frac{a_0 R^2}{4 \nu} \right) \right]
\quad \mbox{and} \quad
\left( u_R, u_Z \right) = \left(- \frac{a_0 R}{2}, a_0Z \right) .
\end{equation}
Here $\nu$ is the kinematic viscosity and $a_0 \ (> 0)$ is a constant. The circulation is maximal at $R$ = $R_0$ = $2.24 (\nu / a_0)^{1/2}$. Thus, $R_0$ is regarded as the tube radius. We do not use other models for vortex tubes, e.g., spirals of \citet{l82}, because detailed information about individual vortex tubes is anyway not available from one-dimensional velocity data.

\begin{figure}[b]
\centering
\begin{minipage}[htbp]{0.3\textwidth}
\begin{center}
\scalebox{0.24}{\includegraphics*[0cm,2.cm][19.5cm,23cm]{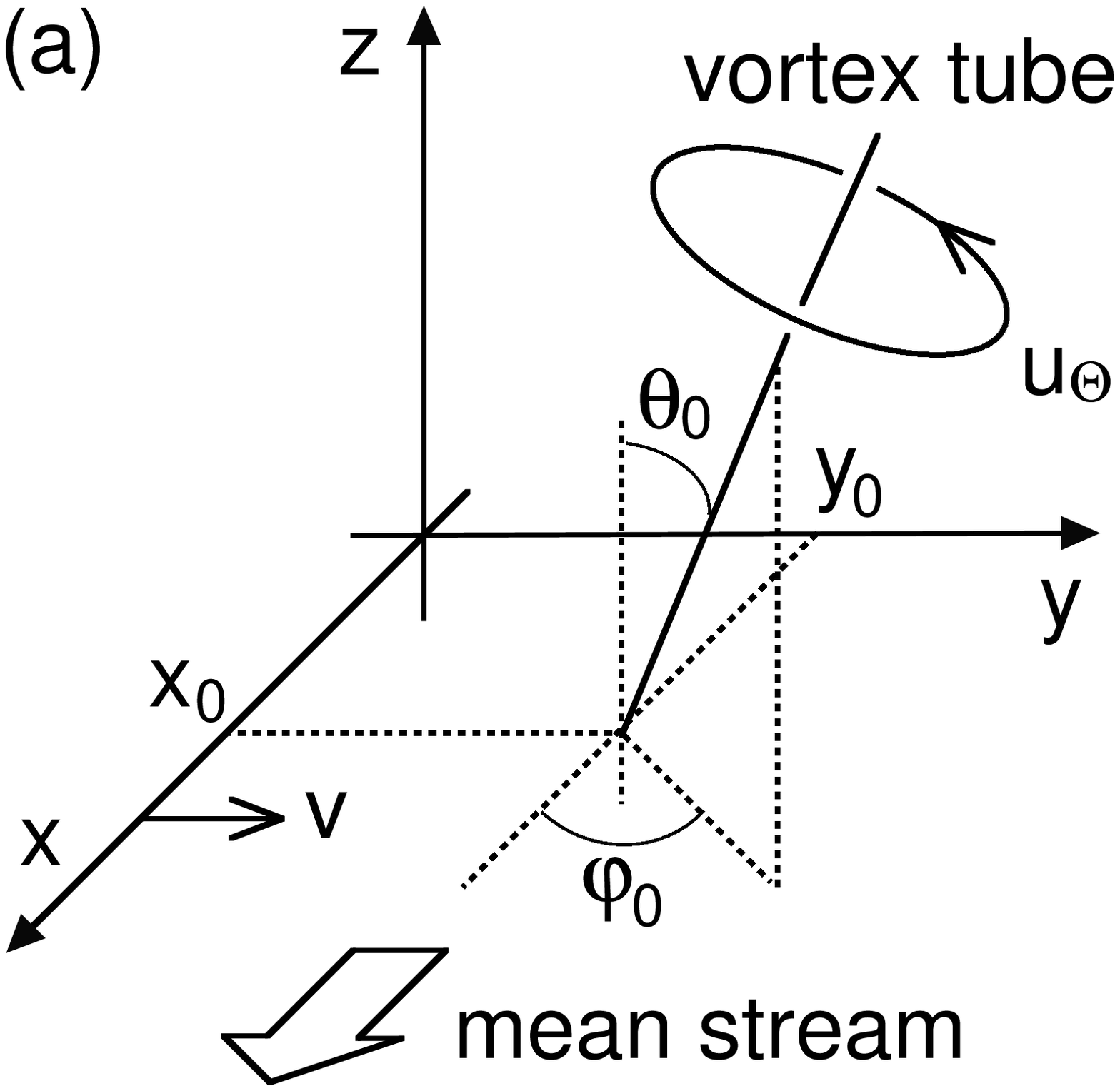}}
\end{center}
\end{minipage}
\hspace{1cm}
\centering
\begin{minipage}[htbp]{0.6\textwidth}
\begin{center}
\scalebox{0.59}{\includegraphics*[4cm,11.6cm][20cm,22.cm]{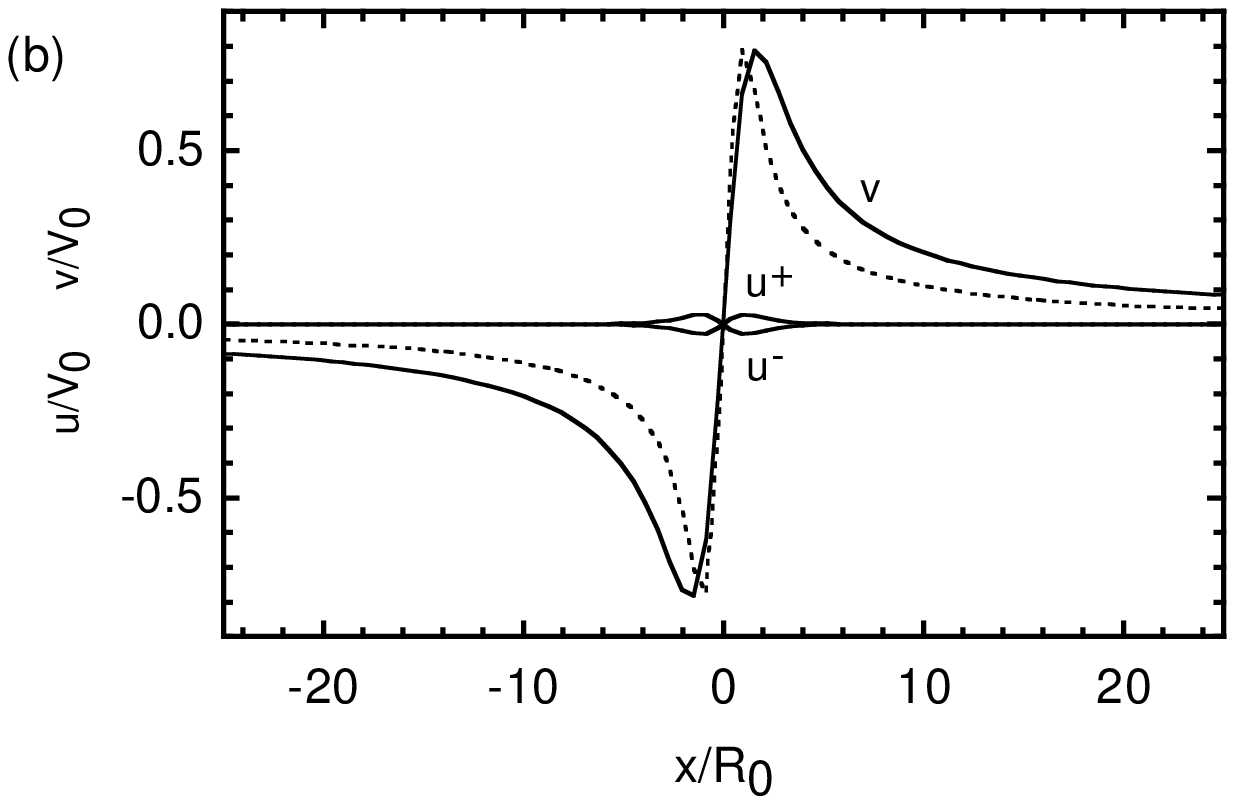}}
\end{center}
\end{minipage}
\caption{\label{f1} (a) Sketch of a vortex tube penetrating the $(x,y)$ plane at a point $(x_0,y_0)$. The inclination is $(\theta_0, \varphi_0)$. The circulation velocity is $u_{\Theta}$. We consider the spanwise velocity $v$ along the $x$ axis in the mean stream direction. (b) Mean profiles for the Burgers vortices with random positions and inclinations. The streamwise velocity $u$ is separately shown for $\partial _x u > 0$ ($u^+$) and $\partial _x u \le 0$ ($u^-$) at $x = 0$. The position and velocities are normalized by the radius $R_0$ and maximum circulation velocity $V_0$ of the Burgers vortices. The dotted line is the $v$ profile of the Burgers vortex for $x_0 = y_0 = \theta_0 = 0$, the peak value of which is scaled to that of the mean $v$ profile.}
\end{figure}

Suppose that velocity data are obtained on a one-dimensional cut of a flow field that consists of vortex tubes and the background random fluctuation, as illustrated in Fig. \ref{f1}a. The one-dimensional cut is along the $x$ axis, the tube position is $(x_0, y_0)$, and the tube inclination is $(\theta _0,\varphi _0)$. The circulation flows $u_{\Theta}$ of the vortex tubes induce small-scale variations in the $v$ signal. 

If we consider intense velocity variations above a high threshold \cite{mhk07}, their scale and amplitude correspond to the radius and circulation velocity of intense vortex tubes with $|y_0| \lesssim R_0$ and $\theta _0 \simeq 0$. To demonstrate this, mean profiles along the cut $x$ are obtained for the circulation flows $u _{\Theta}$ of the Burgers vortices with random positions $(x_0, y_0)$ and inclinations $(\theta _0,\varphi _0)$. Their radii $R_0$ and maximum circulation velocities $V_0 = u_{\Theta}(R_0)$ are set to be the same. We consider the Burgers vortices with $| \partial _x v |$ at $x=0$ being above a threshold, $| \partial _x v |/3$ for $x_0 = y_0 = \theta_0 = 0$ at $x=0$. When $\partial _x v$ is negative, the sign of the $v$ signal is inverted before the averaging. The result is shown in Fig. \ref{f1}b. Around the peaks, the mean $v$ profile is similar to that of the Burgers vortex for $x_0 = y_0 = \theta _0 = 0$ (dotted line).

If we consider all velocity variations that are more significant than the background fluctuation, which roughly correspond to velocity variations above a low threshold, vortex tubes with $|y_0| \gg R_0$ or $\theta _0 \gg 0$ significantly contribute to the mean $v$ profile. Its scale is large while its amplitude is small as compared with the mean radius and circulation velocity of all vortex tubes. Nevertheless, the scale and amplitude of the mean $v$ profile are  proportional to the mean radius and circulation velocity of all vortex tubes, among flow fields where the distribution of vortex tubes is similar. This is likely in our experiments. Since the measured ratio $\langle u^2 \rangle / \langle v^2 \rangle$ is not far from unity (Table \ref{t1}), the tube inclination $(\theta_0,\varphi_0)$ should have been random in all the experiments. Then, we are able to study dependence of those tube parameters on the Reynolds number $Re_{\lambda}$ and on the configuration for turbulence production. 

\begin{figure}[b]
\centering
\begin{minipage}[t]{0.7\textwidth}
\begin{center}
\resizebox{7.3cm}{!}{\includegraphics*[4.5cm,9.5cm][17.5cm,26cm]{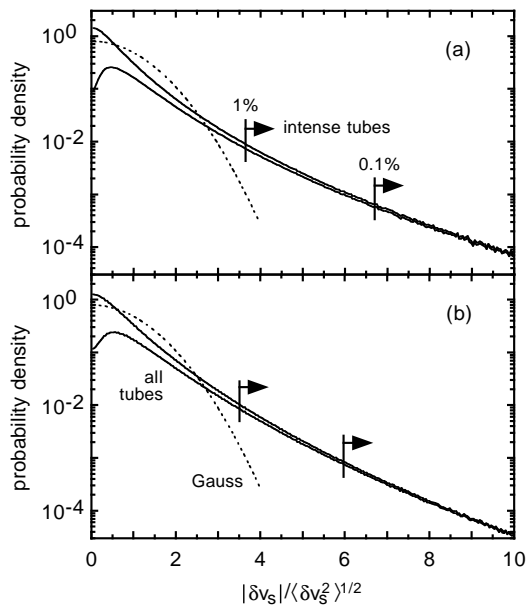}}
\end{center}
\end{minipage}
\caption{\label{f2} Probability density of $|\delta v_s| = \vert v(x+\delta x_s) - v(x) \vert$. (a) $Re_{\lambda} = 719$ (D1). (b) $Re_{\lambda} = 1934$ (D5). We normalize $|\delta v_s|$ by $\langle \delta v_s^2 \rangle ^{1/2} = \langle [v(x+\delta x_s) - v(x)]^2 \rangle ^{1/2}$. The dotted line denotes the Gaussian distribution. The arrows indicate the ranges for intense vortex tubes studied by \citet{mhk07}, which share 0.1 and 1\% of the total. We also show the probability density for all vortex tubes.}
\end{figure}

\section{Mean Velocity Profile} \label{s4}

The spanwise-velocity increment $v(x+ \delta x_0/2) - v(x- \delta x_0/2)$ over a small scale $\delta x_0$ varies at the position $x = x_0$ of a vortex tube, regardless of its intensity. We try to use all of such variations and study vortex tubes as a whole. The increment is smoothed with the Gaussian window function $\exp (-x^2 /2 \delta x_0^2)$. We determine the tube positions $x_0$ as local maxima and minima of the smoothed increment and then analyze the unsmoothed velocity data. The smoothing is to make sure that each of the local maxima and minima corresponds to each vortex tube. Also, the smoothing reduces the background fluctuation. We set $\delta x_0$ to be a multiple of the sampling interval $\delta x_s$ that is close to the mean radius $6\eta$ estimated for intense vortex tubes in \citet{mhk07}. This estimate is consistent with those in \citet{j93}, \citet{b96}, \citet{jw98}, and \citet{tkmsm04}, if we consider the difference in the definition of tube radius.

Fig. \ref{f2} compares the probability density of the absolute velocity increment $\vert v(x+\delta x_s) - v(x) \vert$ over the sampling interval $\delta x_s$ for the entire data with that for the subdata from $x = x_0-\delta x_0$ to $x_0+\delta x_0-\delta x_s$ around all tube positions $x_0$. The latter distribution accounts for the tail of the former distribution where vortex tubes should be dominant \cite{mhk07}, and does not account for the center of the former distribution where the background fluctuation should be dominant. It is thereby expected that we have surely captured vortex tubes.

\begin{figure}[b]
\centering
\begin{minipage}[t]{0.7\textwidth}
\begin{center}
\resizebox{7.3cm}{!}{\includegraphics*[4.5cm,9.5cm][17.5cm,26cm]{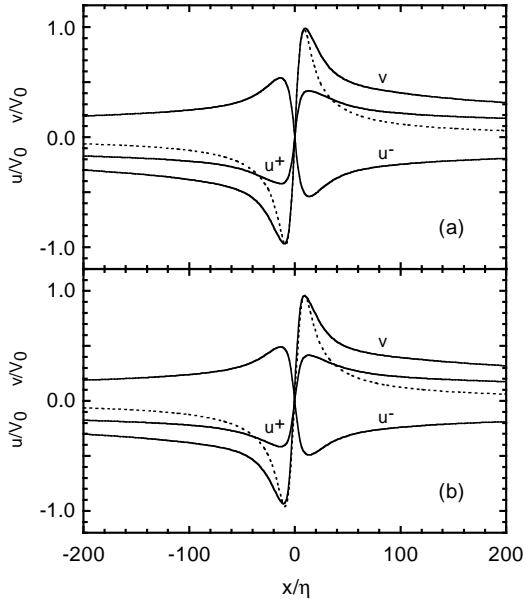}}
\end{center}
\end{minipage}
\caption{\label{f3} Mean profiles of all vortex tubes in the streamwise ($u$) and spanwise ($v$) velocities. (a)  $Re_{\lambda} = 719$ (D1). (b) $Re_{\lambda} = 1098$ (D2). The $u$ profile is separately shown for $\partial _x u > 0$ ($u^+$) and $\partial _x u \le 0$ ($u^-$) at $x = 0$. The dotted line is the $v$ profile fitted by the Burgers vortex for $x_0 = y_0 = \theta_0 = 0$. Its $V_0$ value is used to normalize the velocities. The position $x$ is normalized by the Kolmogorov length $\eta$. For the corresponding figure for intense vortex tubes alone, see Fig. 4 of \citet{mhk07}.}
\end{figure}

Fig. \ref{f3} shows the mean velocity profiles obtained by averaging signals centered at individual tube positions $x_0$. The $v$ profile is similar to that in Fig. \ref{f1}b. Hence, the contribution from vortex tubes is surely dominant. The contribution from vortex sheets is not dominant. If it were dominant, the $v$ profile in Fig. \ref{f3} should exhibit some kind of step \cite{n97,mhk07}. Exceptionally, to the extended tails of the $v$ profile, the contribution from vortex sheets might be dominant.

By fitting the $v$ profile in Fig. \ref{f3} around its peaks by the $v$ profile of the Burgers vortex for $x_0 = y_0 = \theta_0 = 0$ (dotted line), we estimate the radius $R_0$ and maximum circulation velocity $V_0$. The measured velocity is considered to have been smoothed over the probe size in the streamwise direction, 1\,mm. Table \ref{t2} lists the $R_0$ and $V_0$ values. While the $R_0$ value is greater than the true mean radius, the $V_0$ value is less than the true mean circulation velocity, because of the contribution from vortex tubes with $|y_0| \gg R_0$ or $\theta _0 \gg 0$. The $R_0$ and $V_0$ values are still proportional to the true mean values (\S \ref{s3}).

\begin{table}[t]
\begin{center}
\caption{Parameters of vortex tubes in boundary layers (B1--B6) and duct flows (D1--D5): identification scale $\delta x_0$, radius $R_0$, maximum circulation velocity $V_0$, and Reynolds number $Re_0 = R_0 V_0 / \nu$. For parameters of intense vortex tubes alone, see Table 2 of \citet{mhk07}.\label{t2}}
\vspace*{0.4cm}
\begin{tabular}{ccccccc}
\hline
Data   &${\displaystyle \delta x_0 / \eta                }$
       &${\displaystyle R_0 / \eta                       }$
       &${\displaystyle V_0 / u_K                        }$
       &${\displaystyle V_0 / \langle v^2 \rangle ^{1/2} }$
       &${\displaystyle Re_0                             }$
       &${\displaystyle Re_0/Re_{\lambda}^{1/2}          }$ \\
\hline
B1 & 6.30 & 8.45 & 2.43 & 0.263 & 20.5 & 1.13\\
B2 & 5.59 & 8.47 & 2.39 & 0.213 & 20.2 & 0.914\\
B3 & 5.36 & 8.59 & 2.47 & 0.182 & 21.2 & 0.792\\
B4 & 6.06 & 9.21 & 2.73 & 0.175 & 25.1 & 0.817\\
B5 & 6.00 & 8.93 & 2.83 & 0.169 & 25.3 & 0.770\\
B6 & 5.30 & 8.28 & 2.88 & 0.157 & 23.8 & 0.659\\
D1 & 6.16 & 8.56 & 2.43 & 0.179 & 20.8 & 0.776\\
D2 & 5.76 & 8.91 & 2.56 & 0.152 & 22.8 & 0.688\\
D3 & 6.52 & 9.12 & 2.79 & 0.146 & 25.4 & 0.675\\
D4 & 5.84 & 8.72 & 2.91 & 0.138 & 25.4 & 0.617\\
D5 & 6.66 & 8.81 & 3.18 & 0.143 & 28.0 & 0.637\\
\hline
\end{tabular}
\end{center}
\vspace{0.7cm}
\end{table}

The $u$ profile in Fig. \ref{f3} is separated for $\partial _x u > 0$ ($u^+$) and $\partial _x u \le 0$ ($u^-$) at $x = 0$.\footnote{
We have decomposed the $u^{\pm}$ profiles into symmetric and antisymmetric components and show only the antisymmetric components \cite{mhk07}. This is because, although the $u^{\pm}$ profiles for vortex tubes should be antisymmetric, a symmetric positive excursion is induced by the contamination with the $w$ velocity that is perpendicular to the $u$ and $v$ velocities \cite{sm05}. The two wires of the hot-wire anemometer individually respond to all the $u$, $v$, and $w$ velocities. Since the measured $u$ velocity corresponds to the sum of the responses of the two wires, it is contaminated with the $w$ velocity. Since the measured $v$ velocity corresponds to the difference of the responses, it is free from the $w$ velocity.} 
 These $u^{\pm}$ profiles have larger amplitudes than the $u^{\pm}$ profiles in Fig. \ref{f1}b. This is a signature of vortex tubes with $|y_0| \gg R_0$ or $\theta _0 \gg 0$, especially of tubes passing the probe with some incidence angles relative to the mean flow direction, $\tan^{-1} [v/(U+u)]$ \cite{b96}. The radial inflow $u_R$ of the strain field is not discernible, except that the $u^-$ profile has a larger amplitude than the $u^+$ profile. Unlike the Burgers vortex, a real vortex tube is not always oriented to the stretching direction \cite{vm91,vm94,j93,jw98,tkmsm04}.

\begin{figure}[b]
\vspace{0.5cm}
\centering
\begin{minipage}[t]{0.7\textwidth}
\begin{center}
\resizebox{7.3cm}{!}{\includegraphics*[4.5cm,5cm][17.5cm,25cm]{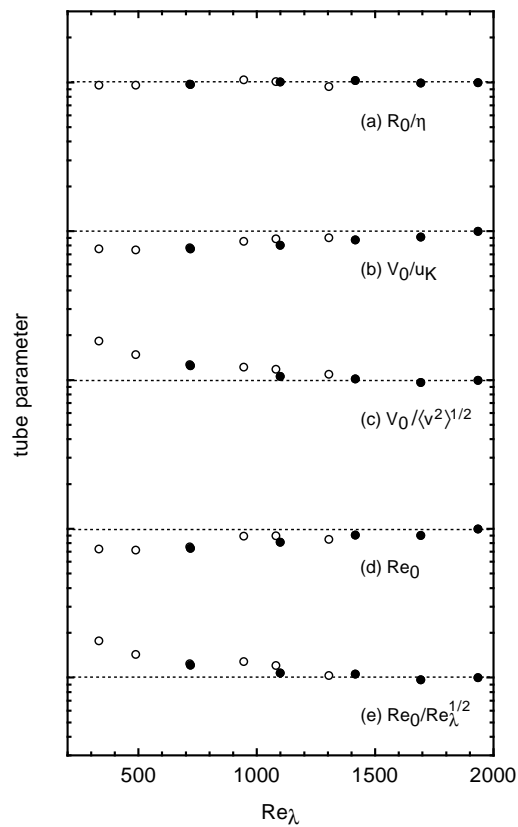}}
\end{center}
\end{minipage}
\caption{\label{f4} Dependence of parameters of all vortex tubes on $Re_{\lambda}$. (a) $R_0/\eta$. (b) $V_0/u_K$. (c) $V_0/\langle v^2 \rangle ^{1/2}$. (d) $Re_0$. (e) $Re_0$/$Re_{\lambda}^{1/2}$. The open circles denote the boundary layers (B1--B6). The filled circles denote the duct flows (D1--D5). Each quantity is normalized by its value in the duct flow at $Re_{\lambda} = 1934$ (D5). For the corresponding figure for intense vortex tubes alone, see Fig. 7 of \citet{mhk07}.}
\end{figure}

\section{Scaling Laws of Tube Parameters} \label{s5}

The dependence of tube parameters on the microscale Reynolds number $Re_{\lambda}$ and on the configuration for turbulence production, i.e., boundary layer or duct flow, is studied in Fig. \ref{f4}. Each quantity was normalized by its value in the duct flow at $Re_{\lambda} = 1934$ (D5). That is, we avoid the prefactors that are not equal to the true values (\S \ref{s4}). We instead focus on scaling laws.

The radius $R_0$ scales with the Kolmogorov length $\eta$ as $R_0 \propto \eta$ (Fig. \ref{f4}a). This is partly because vortex tubes were identified on a scale $\delta x_0 \simeq 6\eta$ (\S \ref{s4}), but the scatter of $R_0/\eta$ is less than the scatter of $\delta x_0/\eta$ as in Table \ref{t2}.

The maximum circulation velocity $V_0$ scales with the Kolmogorov velocity $u_K$ as $V_0 \propto u_K$ (Fig. \ref{f4}b) rather than with the rms velocity fluctuation $\langle v^2 \rangle ^{1/2}$ as $V_0 \propto \langle v^2 \rangle ^{1/2}$ (Fig. \ref{f4}c), if we consider the entire $Re_{\lambda}$ range. This result is reasonable because $u_K$ is a characteristic of small-scale motions. At $Re_{\lambda} \gtrsim 1000$, the scaling $V_0 \propto \langle v^2 \rangle ^{1/2}$ is also significant. Since $V_0 \propto \langle v^2 \rangle ^{1/2}$ is observed for intense vortex tubes (\S \ref{s6}), $V_0 \propto \langle v^2 \rangle ^{1/2}$ is also observed for all vortex tubes if the mean $v$ profile is biased toward intense vortex tubes. This is more likely at higher $Re_{\lambda}$ because $Re_{\lambda} \propto \langle v^2 \rangle/u_K^2$. However, since intense vortex tubes are rare (\S \ref{s6}), $V_0 \propto \langle v^2 \rangle ^{1/2}$ does not represent vortex tubes as a whole. We do not consider the scaling $V_0 \propto \langle v^2 \rangle ^{1/2}$.

Direct numerical simulations of homogeneous isotropic turbulence and turbulent channel flows at $Re_{\lambda} \lesssim 200$ derived the scalings $R_0 \propto \eta$ and $V_0 \propto u_K$ for all vortex tubes \cite{kida,tkmsm04}. We have found that, regardless of the configuration for turbulence production, those scalings extend at least up to $Re_{\lambda} \simeq 2000$.

The scalings of the radius $R_0$ and circulation velocity $V_0$ lead to a scaling of the Reynolds number $Re_0 = R_0 V_0 / \nu$ that characterizes stability of vortex tubes \cite{j93,jw98}:
\begin{subequations} 
\label{eq2}
\begin{eqnarray}
& &Re_0 = \mbox{constant} \quad
{\rm if} \ R_0 \propto \eta \ {\rm and} \
V_0 \propto u_K,  
\label{eq2a} \\
& &Re_0 \propto Re_{\lambda}^{1/2} \quad
{\rm if} \ R_0 \propto \eta \ {\rm and} \
V_0 \propto \langle v^2 \rangle ^{1/2}
\label{eq2b} 
\end{eqnarray}
\end{subequations}
If we consider the entire $Re_{\lambda}$ range, our result favors the former scaling (Fig. \ref{f4}d) rather than the latter (Fig. \ref{f4}e). This is in accordance with the observed scalings of $R_0$ and $V_0$ (Figs. \ref{f4}a and \ref{f4}b). The constancy of $Re_0$ implies that vortex tubes as a whole are stable against an increase of the Reynolds number $Re_{\lambda}$. At $Re_{\lambda} \gtrsim 1000$, the scaling $Re_0 \propto Re_{\lambda}^{1/2}$ is also significant. This scaling is not important to us because it is related with $V_0 \propto \langle v^2 \rangle ^{1/2}$ discussed before (Fig. \ref{f4}c).

\section{Comparison with Intense Vortex Tubes} \label{s6}

The present result for all vortex tubes with various intensities is compared with a past result for intense vortex tubes alone. We use the result of \citet{mhk07}, which was based on the same experimental data. For a similar experimental result for intense vortex tubes, see \citet{b96}.

To identify intense vortex tubes, \citet{mhk07} imposed a threshold on the absolute velocity increment $\vert v(x+\delta x_s) - v(x) \vert$. The threshold was such that 0.1\% or 1\% of the increments were used for the identification. These increments are included in the increments used for all vortex tubes as shown in Fig. \ref{f2}. Thus, intense vortex tubes studied by \citet{mhk07} are included in all vortex tubes studied here.

\citet{mhk07} estimated the radius $R_0$ and maximum circulation velocity $V_0$ from mean velocity profiles. They were obtained by averaging signals centered at individual positions where $\vert v(x+\delta x_s) - v(x) \vert$ was above the threshold. Since the mean velocity profiles were dominated by vortex tubes with $|y_0| \lesssim R_0$ and $\theta _0 \simeq 0$, the $R_0$ and $V_0$ values are close to the true mean radius and true mean circulation velocity (\S \ref{s3}). The dependence of $R_0$ and $V_0$ on the threshold was discussed in \citet{mhk07}.

The radius for intense vortex tubes, $R_0/\eta \simeq 5$--7 \cite{mhk07}, is less than that for all vortex tubes, $R_0/\eta \simeq 8$--9 (Table \ref{t2}). While the former is close to the true mean radius, the latter is greater than the true mean radius (\S \ref{s4}). The true mean radius appears not to be significantly different between intense and all vortex tubes. Both of them obey the scaling $R_0 \propto \eta$. The circulation flows of vortex tubes are always of smallest scales of turbulence.

The circulation velocity for intense vortex tubes, $V_0 / \langle v^2 \rangle^{1/2} \simeq 0.4$--0.8 \cite{mhk07}, is greater than that for all vortex tubes, $V_0 / \langle v^2 \rangle^{1/2} \simeq 0.1$--0.3 (Table \ref{t2}). This is mainly due to the difference in the true mean circulation velocity. In addition, the $V_0$ value for all vortex tubes is less than the true mean circulation velocity (\S \ref{s4}). While all vortex tubes obey the scaling $V_0 \propto u_K$, intense vortex tubes obey the scaling $V_0 \propto \langle v^2 \rangle ^{1/2}$ in the same $Re_{\lambda}$ range.

The Reynolds number for intense vortex tubes scales as $Re_0 \propto Re_{\lambda}^{1/2}$, which is explained through equation (\ref{eq2b}) by the scalings $R_0 \propto \eta$ and $V_0 \propto \langle v^2 \rangle^{1/2}$. With an increase of $Re_{\lambda}$, intense vortex tubes progressively have higher $Re_0$ and are more unstable. The situation is different in the case of all vortex tubes, for which $Re_0$ is constant.

Direct numerical simulations at $Re_{\lambda} \lesssim 200$ derived $R_0 \propto \eta$, $V_0 \propto \langle v^2 \rangle ^{1/2}$, and hence $Re_0 \propto Re_{\lambda}^{1/2}$ for intense vortex tubes \cite{j93,jw98} while $R_0 \propto \eta$, $V_0 \propto u_K$, and hence $Re_0 = \mbox{constant}$ for all vortex tubes \cite{kida,tkmsm04}. This difference in the scalings of $V_0$ and $Re_0$ is the same as that obtained here up to $Re_{\lambda} \simeq 2000$.

The difference in the scalings of $V_0$ and $Re_0$ between intense and all vortex tubes implies that their roles are different. Since $V_0 \propto \langle v^2 \rangle^{1/2}$, intense vortex tubes are responsible for small-scale intermittency. This is especially the case at high $Re_{\lambda}$ because $Re_{\lambda} \propto \langle v^2 \rangle/u_K^2$. In fact, at high $Re_{\lambda}$, small-scale intermittency is significant. However, since $Re_0 \propto Re_{\lambda}^{1/2}$, intense vortex tubes have short lifetimes and thus rare at high $Re_{\lambda}$ \cite{j93,jw98,mhk07}. On the other hand, vortex tubes as a whole obey the scalings $V_0 \propto u_K$ and $Re_0 = \mbox{constant}$. They are always ubiquitous and responsible for an important fraction of energy at smallest scales.

\section{Conclusion} \label{s7}

Using velocity data obtained in boundary layers at $Re_{\lambda} = 332$--1304 and duct flows at $Re_{\lambda} = 719$--1934 \cite{mhk07}, we have studied vortex tubes, i.e., the elementary structures of turbulence. While past experimental studies focused on intense vortex tubes, the present study focuses on all vortex tubes with various intensities. We have obtained the mean velocity profile, estimated the radius $R_0$ and maximum circulation velocity $V_0$, and then obtained the scalings $R_0 \propto \eta$, $V_0 \propto u_K$, and $Re_0 = R_0 V_0/\nu = \rm{constant}$. They are in contrast to the scalings for intense vortex tubes alone, i.e., $R_0 \propto \eta$, $V_0 \propto \langle v^2 \rangle ^{1/2}$, and $Re_0 \propto Re_{\lambda}^{1/2}$. Since those scalings for all vortex tubes are independent of the configuration for turbulence production, they appear to be universal at high Reynolds numbers $Re_{\lambda}$. The implication of the scalings is that vortex tubes as a whole are always ubiquitous and responsible for an important fraction of energy at smallest scales.

The present study has some ambiguities because only one-dimensional data of the velocity field are available. To proceed further, two- or three-dimensional velocity data are necessary but are not available for now. The advent of array of hot-wire probes \cite{sm05} or particle image velocimeter \cite{t02} that is applicable to vortex tubes at high Reynolds numbers is desirable. Such a technique would enlarge our knowledge of vortex tubes. For example, the study of the probability density distributions of the tube radius and maximum circulation velocity is of interest.

\end{document}